\documentstyle[12pt,epsf]{article}
\newcommand{\la}[1]{\label{#1}}
\newcommand{\be}{\begin{equation}}
\newcommand{\ee}{\end{equation}}
\newcommand{\ba}{\begin{eqnarray}}
\newcommand{\ea}{\end{eqnarray}}
\newcommand{\bi}{\begin{itemize}}
\newcommand{\ei}{\end{itemize}}
\newcommand{\rmi}[1]{{\mbox{\scriptsize #1}}}
\newcommand{\fig}{Fig.~}
\newcommand{\eq}{Eq.~}

\newcommand{\nr}[1]{(\ref{#1})}
\newcommand{\tr}{{\rm Tr\,}}

\newcommand{\fr}[2]{{\frac{#1}{#2}}}
\newcommand{\msbar}{\overline{\mbox{\rm MS}}}

\def\lsi{\raise0.3ex\hbox{$<$\kern-0.75em\raise-1.1ex\hbox{$\sim$}}}
\def\gsi{\raise0.3ex\hbox{$>$\kern-0.75em\raise-1.1ex\hbox{$\sim$}}}

\begin{document}

\begin{titlepage}
\begin{flushright}
CERN-TH/99-124\\
hep-lat/9905004 
\end{flushright}
\begin{centering}
\vfill

{\bf THE NON-PERTURBATIVE QCD DEBYE MASS \\ FROM 
A WILSON LINE OPERATOR}
\vspace{0.8cm}

M. Laine$^{\rm a,b,}$\footnote{mikko.laine@cern.ch},
O. Philipsen$^{\rm a,}$\footnote{owe.philipsen@cern.ch}

\vspace{0.3cm}
{\em $^{\rm a}$%
Theory Division, CERN, CH-1211 Geneva 23,
Switzerland\\}
\vspace{0.3cm}
{\em $^{\rm b}$%
Dept.\ of Physics,
P.O.Box 9, FIN-00014 Univ.\ of Helsinki, Finland\\}

\vspace{0.7cm}
{\bf Abstract}

\end{centering}

\vspace{0.3cm}\noindent
According to a proposal by Arnold and Yaffe, the non-perturbative
$g^2T$-contribution to the Debye mass in the deconfined QCD plasma phase
can be determined from a single Wilson line operator in the three-dimensional 
pure SU(3) gauge theory. We extend a previous SU(2) measurement of this 
quantity to the physical SU(3) case. We find a numerical coefficient 
which is more accurate and smaller than that obtained previously with 
another method, but still very large compared with the naive expectation:
the correction is larger than the leading term up to 
$T\sim 10^7 T_c$, corresponding to $g^2\sim 0.4$. 
At moderate temperatures $T\sim 2 T_c$, a consistent picture 
emerges where the Debye mass is $m_D\sim 6T$, 
the lightest gauge invariant screening mass
in the system is $\sim 3T$, and the purely magnetic operators
couple dominantly to a scale $\sim 6T$.
Electric ($\sim gT$) and magnetic ($\sim g^2T$) scales are therefore
strongly overlapping close to the phase transition, and the
colour-electric fields play an essential role in the dynamics. 

\vfill

\noindent
CERN-TH/99-124\\
May 1999

\vfill

\end{titlepage}

\section{Introduction}

An important concept in the description of the deconfined finite
temperature QCD plasma is the Debye mass $m_D$. It is
the inverse of the spatial screening length felt by 
colour-electric excitations, in analogy with Debye screening in 
abelian plasmas, which is felt by electric fields
but not by magnetic fields. 
The leading order result for $m_D$ is perturbatively 
computable~\cite{lo}, but the next-to-leading order result 
is already non-perturbative~\cite{rebhan}, making the
determination of $m_D$ (and, in fact, already 
its precise definition)
a challenging theoretical problem.
The numerical value of $m_D$ has a number of 
applications, in particular in the phenomenology for 
quark-gluon plasma formation (for a review, see~\cite{wang}).
It is also interesting to note that, even though $m_D$ is 
originally a strictly static quantity, it might have 
significance even for some dynamical observables, such as 
the rates of anomalous processes in the plasma~\cite{moore}.

A gauge invariant, non-perturbative definition for $m_D$ 
has been given by Arnold and Yaffe~\cite{ay}. This definition is 
based on a (Euclidian) time reflection operation,
denoted also by ${\cal R}_z$,
which reverses the sign of the temporal 
component of the gauge field: ${\cal R}_z A_0^a\to -A_0^a$
(for a more detailed discussion, see~\cite{ay}). The
Debye mass is defined as
the smallest among the 
exponential fall-offs that are found for gauge invariant operators
odd under this symmetry. A further observation is that due
to finite temperature dimensional 
reduction~\cite{dr}--\cite{adjoint}, 
the measurement of $m_D$
in numerical simulations can be carried out in a
three-dimensional (3d) SU(N) + adjoint
Higgs theory. 
Lattice measurements in this theory were 
performed, both for SU(2) and SU(3), 
in~\cite{mDown}\footnote{Progress towards a lattice measurement 
of $m_D$ directly with 4d lattice simulations has been 
reported in~\cite{dg} (see below). 
Another gauge invariant definition
within the 3d SU(3)+adjoint Higgs model was suggested 
in~\cite{bka}. Other definitions of $m_D$, based on gauge 
fixing, have recently been pursued in~\cite{gf}.}. 

However, there is a problem in making precise simulations 
in the 3d SU(3) + adjoint Higgs theory. The reason is that
in the regime of asymptotically small couplings, there is 
a scale hierarchy between the leading order result $m_D \sim gT$ and the 
non-perturbative correction $\delta m_D \sim g^2T$, which one 
is trying to determine (the ``magnetic'' degrees of freedom
are also associated with the scale $g^2T$). 
As always, a scale hierarchy is difficult to accommodate on a single
lattice of presently available sizes, and thus makes 
the approach to the continuum limit quite
problematic in practical simulations. 

Fortunately, there is a way out:
parametrically (i.e., as a power series in $g$), 
even the field $A_0^a$ can be integrated
out perturbatively, leaving a pure 3d SU(3) gauge theory. 
The operator odd in $A_0^a$ goes over to a non-local
but gauge invariant Wilson line
in the pure SU(3) gauge theory~\cite{ay}. 
This allows to determine the coefficient of $g^2T$.

For the SU(2) case, this Wilson line operator was 
studied with lattice simulations in~\cite{ffsu2}. We found 
a result for the $g^2T$ term
which is more precise and smaller than 
that found in~\cite{mDown}. The purpose of the present
paper is to extend the SU(2) measurement to the physical
SU(3) case.

\section{The operator in the continuum}

Let us start by briefly reviewing 
the basic idea of the approach in~\cite{ay}.
Consider gauge invariant operators $O(x)$ 
odd under $A_0\to -A_0$
(i.e., in the notation of~\cite{ay}, with ${\cal R}_z=-1$). 
In practice, we consider $O(x)$ containing only one power of $A_0$,
since otherwise (for $A_0^n, n\ge3$)
the leading behaviour for small $g$
is expected to be of the form
$n \cdot m_0$, where 
$m_0=(N/3+N_f/6)^{1/2}gT$
is the leading order Debye mass~\cite{lo}
($N_f$ is the number of flavours).  
Imagine now doing the path integral over $A_0$ before doing that over $A_i$. 
Then, 
\be \label{a0}
\frac{\int\! {\cal D} A_0\, O(x) O(y) \exp(-S_{A_0})}
{\int\! {\cal D} A_0\, \exp(-S_{A_0})} \sim
O'_a(x) G^{ab}(x,y) O'_b(y),
\ee
where $O' = d O/d A_0$, 
\be
S_{A_0} = \int\! d^3 x \,\Bigl[
\fr12 (D_i A_0)^a (D_i A_0)^a + \fr12 m_0^2 A_0^a A_0^a
\Bigr],
\ee
and $G^{ab}(x,y)$ 
is the tree-level Green's function in 
the background of $A_i$:
\be
G^{ab}(x,y) = \frac{\exp(-m_0 |x-y|)}{4\pi |x-y|} 
W^{ab}(x,y), \la{pert}
\ee
with $W^{ab}(x,y)$ the adjoint Wilson line,
\be
W^{ab}(x,y) = 
2 \tr T^a W^{\rm fund}(x,y)T^b [W^{\rm fund}(x,y)]^\dagger,
\ee
where  $W^{\rm fund}(x,y) =  {\cal P} 
\exp({ig\int_x^y\! dx_i A_i^a T^a})$.

Thus, the measurement of the Debye mass can be reduced to 
the measurement of 
\ba
G_{F}(x,y) & \equiv  & 
\left\langle O_a'(x) W^{ab}(x,y) O_b'(y)
\right\rangle, 
\la{GA}
\ea
in the pure SU(N) gauge theory. The measurement of this operator
produces an exponential decay with a coefficient $m$, to which 
the perturbative part $m_0$
from \eq\nr{pert} has to be added in order
to obtain the physical result. 
 
Including also loop corrections (i.e., allowing $A_i$ to fluctuate at
all length scales), maintains the form of the answer but replaces the
perturbative $m_0$ in \eq\nr{pert}
by an effective parameter $m_0^\rmi{eff}$~\cite{ay}:
\be
m_0 \to m_0^\rmi{eff} = 
m_0 +  \frac{N g_3^2}{4\pi} \biggl(\ln\frac{m_0}{\Lambda} + 
d_\Lambda \biggr)  + {\cal O}\biggl(g_3^2 \frac{m_0}{\Lambda}, 
\frac{g_3^4}{m_0},...\biggr), 
\ee
where $g_3^2=g^2 T$, 
$\Lambda$ is some ultraviolet cutoff, 
and $d_\Lambda$ depends on $\Lambda$. In particular,  
in the case of a lattice regularization, 
$\Lambda=a^{-1}$ and $d_\Lambda=-(3\ln 2+1)/2$ 
(see \cite{ay,ffsu2}).
It follows that if the physical 
Debye mass is written as
\be
m_D = m_0+{Ng_3^2\over4\pi}\ln{m_0\over  
g_3^2} +  
c_N g_3^2 + {\cal O}(g^3T),
\label{md4d}
\ee
then
\be
c_N =\frac{m}{g_3^2}+             
\frac{N}{8\pi}\left(\ln\frac{N^2}{2\beta_G^2}-1\right),
\ee
where $m$ is the mass measured from the operator
in \eq\nr{GA}, and $\beta_G=2 N/(ag_3^2)$.

We thus need to find operators $O'$ such that we 
get the lowest possible coefficient for the exponential 
falloff in \eq\nr{GA}. 
{}From~\cite{mDown}, we expect the operators for \eq\nr{a0} to be 
$O \sim \epsilon_{3jk} \tr A_0 F_{jk}$ and 
suitable variants thereof (see below), leading to
$O'_a =F^a_{jk}$ in \eq\nr{GA}. In principle, 
one could get rid of the specification of the operators
altogether by measuring the coefficient of the perimeter law 
of the adjoint Wilson loop~\cite{ay}. However, in practice
the perimeter law can only be observed on the lattice by 
including this very kind of operators in the basis, due
to a very small overlap of the lattice Wilson loop with
the asymptotic lightest state~\cite{stringbreak}.

\section{A lattice formulation}

On the lattice, 
we replace \eq\nr{GA} with operators of the type
\be \label{wadj}
G_{F,ijkl}(x,y)=
\left\langle
4 \tr ( F_{ij}(x) T^a) 
\Gamma^{ab}(x,y) 
\tr ( F_{kl}(y)T^b)\right\rangle, 
\la{GF}
\ee 
where $F_{ij}$ is the naive lattice analogue of the field strength tensor, 
\be
F_{ij}(x) = -\frac{i}{2} \Bigl[
P_{ij}(x) - P^\dagger_{ij}(x)
\Bigr],
\ee
$P_{ij}$ is the plaquette, 
and $\Gamma^{ab}$ is the lattice Wilson line:
\ba
\Gamma^{ab}(x,y) & = & 2\tr 
\left (T^a S(x,y) T^b S^{\dag}(x,y) \right),
\\
S(x,y) & = & \prod_{n=0}^{N-1} U_{3}(x+n\hat{3}), 
\qquad y=x+N\cdot \hat{3}. 
\la{sdef}
\ea
Here $U_i$ is a link variable and
$\hat{3}$ is the unit vector in the $x_3$-direction. 
\eq\nr{GF} may be rewritten as
\ba
G_{F,ijkl}(x,y) & = & 2 \left \langle 
\tr \left [ F_{ij}(x)S F_{kl}(y)S^{\dag} 
\right ] - \frac{1}{N} \tr F_{ij}(x) \tr F_{kl}(y) \right \rangle,
\ea
where the latter term vanishes fast close to the 
continuum limit and is not numerically important
(for SU(2), it vanishes identically). 
In our simulations,
we use $G_F$ in this last form, with the components $ij=kl=12$.

In our practical simulations,
each plaquette at the end of the Wilson line 
was replaced by the sum over all four spatial 
plaquettes of the same orientation sharing
the end point of the Wilson line, a configuration 
called the ``clover". This operator is expected to have a better
projection onto the ground state (i.e.~to be less contaminated
by higher spin excitations \cite{mt}), as well as to improve 
on statistics.
We have carried out measurements 
using single plaquettes, as well, and checked that the two operator types
give consistent results, with the clover
indeed providing a much better signal.
In order to further increase the projection onto the lightest mass, we
used smearing and diagonalization, as in~\cite{ffsu2}. The idea
is to measure a whole cross correlation matrix
within a channel of given quantum numbers, and then to 
search for the optimal operator by a variational calculation. The basis
is obtained from gauge invariant ``smeared'' operators with 
different numbers of smearing iterations. 
For details, we refer to~\cite{ffsu2}.

\begin{figure}[t]

\vspace*{-2cm}

\centerline{\epsfxsize=12cm\hspace*{-1cm}\epsfbox{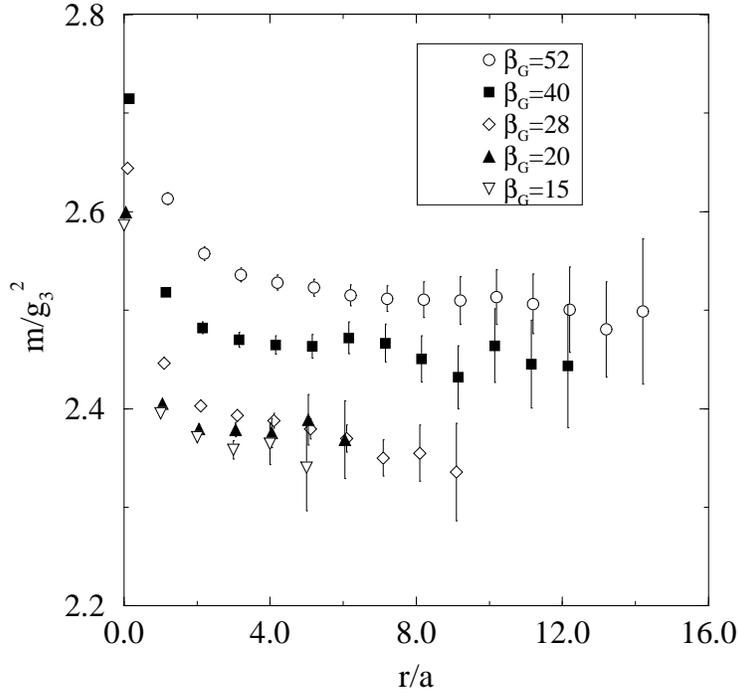}}

\vspace*{-5.5cm}

\caption[a]{The local mass values, $m\,a =-\ln [G_F(n+1)/G_F(n)]$,
from typical diagonalized operators at the different $\beta_G$'s. 
The masses have been converted to continuum units
by multiplying the lattice values $m\, a$ with $\beta_G/6$.}  
\la{fig:timeslices}
\end{figure}

We gathered statistics until the statistical errors of the local
masses at their plateaux were well below 10\%, which required between
500 and 4000 measurements, depending on the lattice size.
The choice of lattice sizes was governed by the experience from
SU(3) glueball measurements~\cite{mt} as well as our 
previous SU(2) measurements~\cite{ffsu2}.
We have also explicitly checked that finite volume 
effects are within the statistical errors at single values of $\beta_G$
(see Tables~\ref{table:su2}, \ref{table:su3}), 
and then increased the volume with $\beta_G$, 
keeping it fixed in physical units.

\section{Numerical results}

Typical local mass values from the correlation 
functions in the case of SU(3) are shown in \fig\ref{fig:timeslices},
exhibiting rather pronounced plateaux.
The mass values quoted in this paper have been obtained 
by fitting an exponential to the optimal diagonalized 
correlation function, and choosing among various fitting ranges those
giving the best values for $\chi^2/{\rm dof}$. 
The lengths of the fitting ranges are between 5 and 11 timeslices, 
and the $\chi^2$-values are between 0.5 -- 1.5.
We now discuss our results both for SU(2) and SU(3).

\vspace*{0.3cm}

{\bf SU(2)}. Previously we carried out measurements
for SU(2) with $\beta_G=9,12,16$~\cite{ffsu2}. We have now 
added values closer to the continuum limit, $\beta_G=20,25$.
The results are shown in Table~\ref{table:su2} and in 
\fig\ref{fig:su2}. As a continuum extrapolation we 
obtain $c_2 = 1.14(4)$. Note that 
we see the onset of a quadratic ${\cal O}(a^2)$
term in $c_2$ on coarser lattices
(the same is true also for SU(3)). 

\begin{table}[t]
\centering
\begin{tabular}{lllll}
\hline
$\beta_G$ & volume         & $m\, a$  & $m/g_3^2$ & $c_2$ \\ \hline
       9  & $30^3$         & 0.685(8) & 1.54(2) & 1.17(2) \\
       12 & $30^2\cdot 42$ & 0.517(7) & 1.55(2) & 1.13(2) \\
       16 & $42^3$         & 0.401(4) & 1.60(2) & 1.14(2) \\
          & $54^3$         & 0.397(3) & 1.59(2) & 1.12(2) \\
       20 & $52^3$         & 0.327(5) & 1.64(3) & 1.13(3) \\
       25 & $64^3$         & 0.268(2) & 1.68(2) & 1.14(2) \\
$\infty$  &                &  --      &   --    & 1.14(4) \\ \hline
\end{tabular}
\caption[a]{\protect 
The results for SU(2). 
The continuum extrapolation is from a linear 
fit to $\beta_G\ge 12$, with $\chi^2$/d.o.f.\ $=0.03$
(see \fig\ref{fig:su2}).}
\la{table:su2}
\end{table}

\begin{figure}[t]


\centerline{\epsfxsize=8cm\hspace*{-1cm}\epsfbox{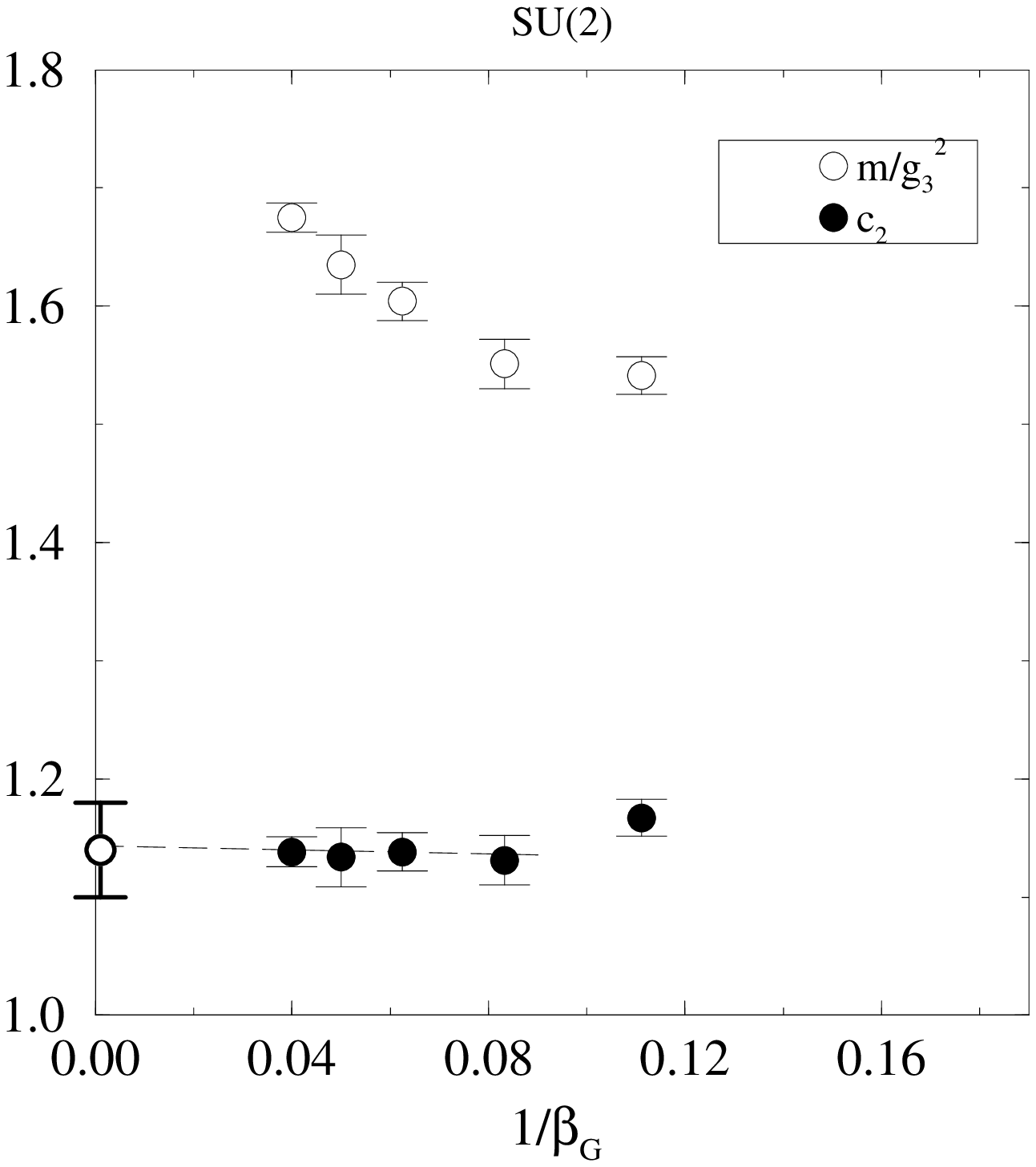}%
\epsfxsize=8cm\hspace*{-1cm}\epsfbox{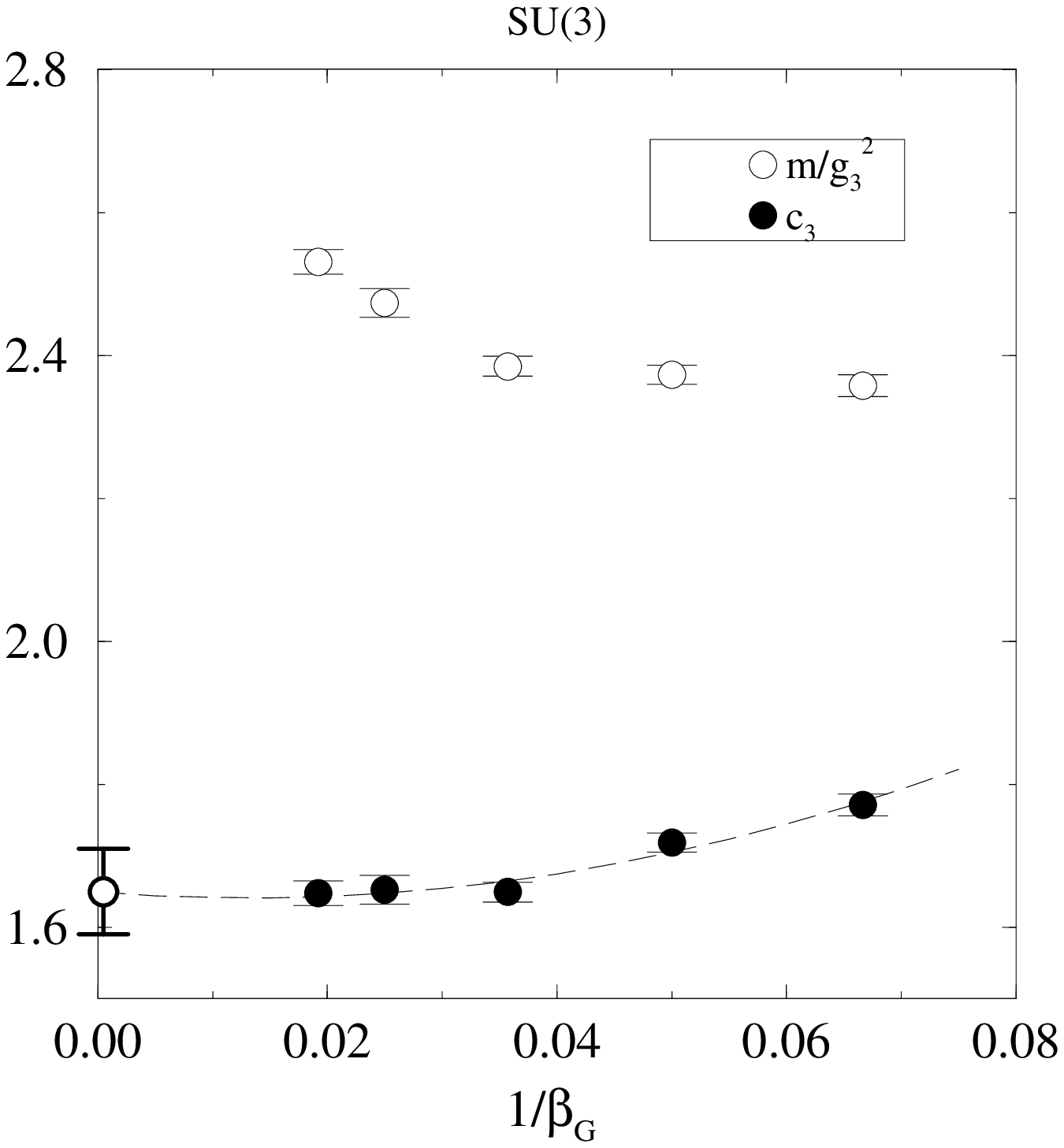}}

\vspace*{-3cm}

\caption[a]{The results for SU(2) (left)
and SU(3) (right). The continuum fits from Tables~\ref{table:su2},
\ref{table:su3} have also been shown. The ranges of $1/\beta_G$
on the $x$-axes have
been chosen such that in lattice units, 
the masses have comparable values for SU(2) and SU(3).} 
\la{fig:su2}
\end{figure}

\vspace*{0.3cm}

{\bf SU(3)}. The results for SU(3) are shown in Table~\ref{table:su3}
and in \fig\ref{fig:su2}. As for SU(2), the results show could scaling
with $\beta_G$ after the removal of the logarithmic divergence. As a 
final continuum value we cite $c_3 = 1.65(6)$.

\begin{table}[t]
\centering
\begin{tabular}{lllll}
\hline
$\beta_G$ & volume         & $m\, a$  & $m/g_3^2$ & $c_3$ \\ \hline
       15 & $18^3$         & 0.943(6) & 2.36(2) & 1.77(2) \\
       20 & $24^3$         & 0.712(4) & 2.37(2) & 1.72(2) \\
          & $32^3$         & 0.707(5) & 2.36(2) & 1.70(2) \\
       28 & $32^3$         & 0.511(3) & 2.38(2) & 1.65(2) \\
       40 & $46^3$         & 0.371(3) & 2.47(2) & 1.65(2) \\
       52 & $60^3$         & 0.292(2) & 2.53(2) & 1.65(2)\\
$\infty$  &                &  --      &   --      & 1.65(6) \\ \hline
\end{tabular}
\caption[a]{\protect 
The results for SU(3). The continuum extrapolation is from a quadratic 
fit to all the $\beta_G$-values, with $\chi^2$/d.o.f.\ $=1.25$. We use
a quadratic fit for SU(3) since we are slightly farther from the 
continuum limit than for SU(2), see \fig\ref{fig:su2}. A linear
fit to $\beta_G\ge28$ gives 1.65(5), with $\chi^2$/d.o.f.\ $=0.04$.}
\la{table:su3}
\end{table}

\section{Discussion}

{\bf Comparison with other measurements in 3d.}
Let us compare our results with those of previous studies.
In~\cite{mDown}, $c_2,c_3$ were measured using operators 
$O \sim \epsilon_{3jk} \tr A_0 F_{jk}$ in the 3d 
SU(N) + adjoint Higgs theory, 
with the result $c_2=1.58(20)$, 
$c_3=2.46(15)$. In~\cite{ffsu2},  
we found the smaller result $c_2=1.06(4)$,
using the field strength correlator 
in the pure 3d SU(2) theory
as in this work. 
Adding two more lattice spacings
to SU(2) and evaluating also SU(3), our present results are
$c_2=$1.14(4), $c_3=$1.65(6). 
In conclusion, the values are considerably
smaller than in~\cite{mDown}. However, there 
the system had multiple length scales, so it need not
be a surprise that the result has some systematic 
uncertainties. Our present results scale
quite well with $N$, a behaviour which is also observed for the glueball
spectrum in SU(N) theories \cite{mt}.
Furthermore, the bare numbers $m/g_3^2$ extracted 
from the Wilson lines exhibit the logarithmic divergence 
expected from lattice perturbation theory \cite{ay}, 
adding further confidence to them. 

\vspace*{0.3cm}

{\bf The mass scales in units of temperature.}
We now proceed to estimate the numerical value
of the Debye mass $m_D$ in units of $T$. For definiteness
we take $N_f=0$, in which case 
$T_c/\Lambda_\rmi{$\msbar$}\approx 1.03(19)$
for SU(3)~\cite{tc}.
(Let us stress that the values of $c_2,c_3$ as such do not depend on 
the fermionic content of the original 4d theory, 
and thus $c_3$ applies also to the physical QCD
with $N_f\neq 0$.)
At moderate temperatures of, say, $2T_c$, we then estimate that 
$g_3^2\approx 2.7T$~\cite{adjoint}, and thus $m_D\sim 6T$. 
This estimate is of course somewhat rough, since terms
of order $g^3T$ etc.\ have been neglected\footnote{Note also
that in principle, some other ${\cal R}_z=-1$ 
operator such as $\tr A_0^3$ could have a lighter 
mass in this region.}. However, the results 
in~\cite{mDown} suggest that these corrections should 
come with very small coefficients. 

An important question for the construction of effective theories 
concerns the degree to which the 
Debye mass is perturbative, and the related issue of how well the
electric ($\sim gT$) and magnetic ($\sim g^2T$) 
sectors are separated, once non-perturbative coefficients are taken into
account.
Even at very large temperatures, 
say,  $T\sim 10^4 T_c$ with a weaker coupling
$g_3^2 \approx 0.64T$, the correction ($\sim 1.1T$)
is still larger than the leading term ($\sim 0.8T$).
Both are of the same order ($\sim 0.7T$) only at $T\sim 10^7 T_c$.

It is also interesting to compare with the masses 
in the magnetic sector, especially at low temperatures 
$T\sim 2T_c$. There the ordering of states according to the 
naive expectations is reversed. 
A purely magnetic excitation, the lightest glueball, has a mass 
$\sim 2.4 g_3^2\sim 6T$~\cite{mt}, making it as heavy as
$m_D$. However, there is a lighter excitation 
with the same quantum numbers which then determines, e.g., 
the decay of the Polyakov loop. In the 3d SU(3) + adjoint 
Higgs theory, it is represented by the mass of the 
bound state $A_0^aA_0^a$, whose numerical value
is about $\sim 3T$ ~\cite{adjoint,suNown}.

\vspace*{0.3cm}

{\bf Comparison with measurements in 4d.}
All these numbers are in  
good qualitative agreement with those found with direct 4d simulations
in~\cite{dg}. In Table 5 of~\cite{dg}, $m(A_1^{++})\approx 2.60(4)T$ 
corresponds to the lightest mass in the $0^{++}$ channel
[in the language of the 3d theory, to $m(A_0^aA_0^a)\sim 3T$], while 
the ${\cal R}_z=-1$ \cite{ay,dg} mass 
$m(A_1^{-+})\approx 6.3(2)T$ should correspond
to the lightest mass in the electric sector
[in the language of the 3d theory, to $m_D$ as defined here, $\sim 6T$].
Since the 4d measurement of $m_D$ contains all orders in $g$ this
agreement would indeed suggest corrections of order $g^3T$
and higher to be small. 

Incidentally, we believe that,  
contrary to the suggestion in~\cite{dg},
the lightest masses measured in~\cite{dg} 
do not correspond to the glueball masses measured
in~\cite{mt}, but to scalar states involving $A_0$: 
the screening masses corresponding to 
static glueball correlators are heavier.
Thus, close to $T_c$, the naive expectation 
that $A_0$ is heavy and decouples is completely 
unjustified (see also~\cite{bn,adjoint,reisz}). 
This should not be a surprise, since
the actual transition is assumed to be driven by
the Z($N$) symmetry related to the Polyakov line and $A_0$. 
At the same time, it is interesting to note that the higher 
lying glueball spectrum itself is expected to be relatively
little affected by $A_0$~\cite{ptw}.

\vspace*{0.5cm}

{\bf Conclusions.}
In conclusion, we have determined the non-perturbative 
contribution of the soft modes $\sim g^2T$ to the Debye mass
in QCD and found it to be larger than the ``leading" perturbative 
contributions for all reasonable temperatures. The approximate 
agreement between the values for the Debye mass as determined 
in this simulation and with the 4d measurement, further suggests 
that the ${\cal O}(g^3T)$ and higher corrections in the Debye mass 
are small even at relatively small temperatures, 
as already argued in~\cite{mDown}. The consistent
picture emerging from our results, as well as from previous
analytical work and simulations~\cite{bn,adjoint,mDown,reisz}, 
is that a dimensionally reduced theory gives a reliable description 
of correlation functions down to quite 
moderate temperatures $T\sim 2T_c$, when $A_0$ is kept
in the action. However, at such temperatures $A_0$ may not be integrated 
out, but is in fact an essential constituent in the lightest physical
degrees of freedom. 

\section*{Acknowledgements}

We thank M. Teper for providing us with a 3d SU(3) 
Monte Carlo code, and for useful discussions. The simulations 
were carried out with a Cray C94 at the Center 
for Scientific Computing, Finland, and with a 
NEC-SX4/32 at the HLRS Universit\"at Stuttgart. This work was 
partly supported by the TMR network {\em Finite Temperature Phase
Transitions in Particle Physics}, EU contract no.\ FMRX-CT97-0122.



\begin{thebibliography}{99}

\bibitem{lo}
E. Shuryak, Zh. Eksp. Teor. Fiz. 74 (1978) 408
[Sov. Phys. JETP 47 (1978) 212];
J. Kapusta, Nucl. Phys. B 148 (1979) 461;
D. Gross, R. Pisarski and L. Yaffe, 
Rev. Mod. Phys. 53 (1981) 43.

\bibitem{rebhan} A.K. Rebhan,
Phys. Rev. D 48 (1993) R3967 [hep-ph/9308232];
Nucl. Phys. B 430 (1994) 319 [hep-ph/9408262].

\bibitem{wang}
X.-N.\ Wang, 
Phys.\ Rep.\ 280 (1997) 287 [hep-ph/9605214].

\bibitem{moore}
G.D. Moore, hep-ph/9810313.

\bibitem{ay}
P. Arnold and L. Yaffe, 
Phys.\ Rev.\ D 52 (1995) 7208 [hep-ph/9508280].

\bibitem{dr}
P.~Ginsparg,
Nucl.\ Phys.\ B 170 (1980) 388;
T. Appelquist and R. Pisarski,
Phys.\ Rev.\ D 23 (1981) 2305.

\bibitem{generic} 
K. Kajantie, M. Laine, K. Rummukainen and M. Shaposhnikov,
Nucl.\ Phys.\ B 458 (1996) 90 [hep-ph/9508379];
Phys.\ Lett.\ B 423 (1998) 137 [hep-ph/9710538].

\bibitem{bn}
E. Braaten and A. Nieto,
Phys.\ Rev.\ Lett.\ 76 (1996) 1417 [hep-ph/9508406];
Phys.\ Rev.\ D 53 (1996) 3421 [hep-ph/9510408];
A. Nieto, 
Int.\ J.\ Mod.\ Phys.\ A 12 (1997) 1431 [hep-ph/9612291].

\bibitem{adjoint}
K. Kajantie, M. Laine, K. Rummukainen and M. Shaposhnikov, 
Nucl.\ Phys.\ B 503 (1997) 357 [hep-ph/9704416].

\bibitem{mDown}
K. Kajantie, M. Laine, J. Peisa, A. Rajantie, K. Rummukainen and
M. Shaposhnikov, 
Phys.\ Rev.\ Lett.\ 79 (1997) 3130 [hep-ph/9708207].

\bibitem{dg}
S. Datta and S. Gupta, 
Nucl.\ Phys.\ B 534 (1998) 392 [hep-lat/9806034].

\bibitem{bka}
S. Bronoff and C.P. Korthals Altes, 
talk at Lattice '98, Boulder, Colorado, 1998
[hep-lat/9808042];
Ph. Boucaud and C.P. Korthals Altes, 
talk at SEWM '98, Copenhagen, Denmark, 1998 [hep-lat/9904006].

\bibitem{gf}
U.M. Heller, F. Karsch and J. Rank,
Phys.\ Rev.\ D 57 (1998) 1438 [hep-lat/9710033];
A. Patk\'os, P. Petreczky and Z. Szep, 
Eur.\ Phys.\ J.\ C 5 (1998) 337 [hep-ph/9711263];
F. Karsch, M. Oevers and P. Petreczky, 
Phys.\ Lett.\ B 442 (1998) 291 [hep-lat/9807035].

\bibitem{ffsu2}
M. Laine and O. Philipsen, 
Nucl.\ Phys.\ B 523 (1998) 267 [hep-lat/9711022].

\bibitem{stringbreak}
O. Philipsen and H. Wittig, 
Phys.\ Rev.\ Lett.\ 81 (1998) 4056 [hep-lat/9807020];
Phys.\ Lett.\ B 451 (1999) 146 [hep-lat/9902003];  
F. Knechtli and R. Sommer, 
Phys.\ Lett.\ B 440 (1998) 345 [hep-lat/9807022];
P.~Stephenson, hep-lat/9902002.

\bibitem{mt}
M. Teper, 
Phys.\ Rev.\ D 59 (1999) 014512 [hep-lat/9804008].

\bibitem{tc}
J. Fingberg, U. Heller and F. Karsch, 
Nucl.\ Phys.\ B 392 (1993) 493 [hep-lat/9208012].

\bibitem{suNown}
K. Kajantie, M. Laine, A. Rajantie, K. Rummukainen and M. Tsypin, 
JHEP 11 (1998) 011 [hep-lat/9811004].

\bibitem{ptw}
O. Philipsen, M. Teper and H. Wittig,
Nucl.\ Phys.\ B 469 (1996) 445 [hep-lat/9602006];
Nucl.\ Phys.\ B 528 (1998) 379 [hep-lat/9709145];
E.M. Ilgenfritz, A. Schiller and C. Strecha,
Eur.\ Phys.\ J.\ C 8 (1999) 135 [hep-lat/9807023];
O. Philipsen, 
talk at SEWM '98, Copenhagen, Denmark, 1998 [hep-ph/9902376].

\bibitem{reisz}
T. Reisz, Z.\ Phys.\ C 53 (1992) 169;
L. K\"arkk\"ainen, P. Lacock, D.E. Miller, B. Petersson 
and T. Reisz, 
Phys.\ Lett.\ B 282 (1992) 121;
Nucl.\ Phys.\ B 418 (1994) 3 [hep-lat/9310014];
L. K\"arkk\"ainen, P. Lacock, B. Petersson
and T. Reisz,
Nucl.\ Phys.\ B 395 (1993) 733.

\end{thebibliography}
\end{document}